\def\MPL #1 #2 #3 {Mod.~Phys.~Lett.~{\bf#1},\  #2 (#3)}
\def\NPB #1 #2 #3 {Nucl.~Phys.~{\bf#1},\  #2 (#3)}
\def\PLB #1 #2 #3 {Phys.~Lett.~{\bf#1},\  #2 (#3)}
\def\PR #1 #2 #3 {Phys.~Rep.~{\bf#1},\ #2 (#3)}
\def\PRD #1 #2 #3 {Phys.~Rev.~{\bf#1},\  #2 (#3)}
\def\PRL #1 #2 #3 {Phys.~Rev.~Lett.~{\bf#1},\  #2 (#3)}
\def\RMP #1 #2 #3 {Rev.~Mod.~Phys.~{\bf#1},\  #2 (#3)}
\def\ZP #1 #2 #3 {Z.~Phys.~{\bf#1},\  #2 (#3)}
\def\IJMP #1 #2 #3 {Int.~J.~Mod.~Phys.~{\bf#1},\  #2 (#3)}

\def\chisq{\chi^2}
\def\ltot{L_{\rm total}}
\def\epst{\eps(t\anti t\h)}
\def\epszh{\eps(Z\h)}
\def\lefft{L_{\rm eff}(t\anti t\h)}
\def\leffzh{L_{\rm eff}(Z\h)}
\def\sigmat{\sigma_T(t\anti t\h)}
\def\sigmazh{\sigma_T(Z\h)}
\def\thdm{2HDM}
\def\anti{\overline}

\def\rts{\sqrt s}
\def\eps{\epsilon}
\def\h{h}
\def\mh{m_{\h}}

\def\epem{e^+e^-}
\def\mupmum{\mu^+\mu^-}

\def\lsim{\mathrel{\raise.3ex\hbox{$<$\kern-.75em\lower1ex\hbox{$\sim$}}}}
\def\gsim{\mathrel{\raise.3ex\hbox{$>$\kern-.75em\lower1ex\hbox{$\sim$}}}}
\def\@versim#1#2{\vcenter{\offinterlineskip
        \ialign{$\m@th#1\hfil##\hfil$\crcr#2\crcr\sim\crcr } }}

\def\ie{{\em i.e.}}

\def\anti{\overline}

\def\fbi{~{\rm fb}^{-1}}
\def\fb{~{\rm fb}}

\def\gev{\,{\rm GeV}}
\def\tev{\,{\rm TeV}}

\def\tanb{\tan\beta}

\def\mt{m_t}
\def\mb{m_b}

\def\h{h}
\def\mh{m_{\h}}

\documentclass[smus]{snow2e}
\input psfig.sty

\begin{document}
\title{
{\normalsize UCD-96-23 \hspace*{\fill} September, 1996} \\
Discriminating between Higgs Boson models using 
$e^+e^-\to t \anti t \h$ and $Z\h$ at the NLC
\thanks{
To appear in ``Proceedings of the 1996 DPF/DPB Summer Study
on New Directions for High Energy Physics''.
Work supported in part by the Department of Energy,
 by the Davis Institute for High Energy Physics and by
the Australian Research Council.}}

\author{J.~F. Gunion (U.C. Davis) and X.-G. He (Melbourne)\\ 
}

\maketitle

\thispagestyle{empty}\pagestyle{empty}

\begin{abstract}
We demonstrate that the process $e^+e^-\to t\anti t h$ at the 
NLC provides a powerful tool for extracting the $t\anti t$ (Yukawa)
couplings of the $\h$. In combination with the $\epem\to Z\h$ process,
an accurate determination of the $ZZ$ coupling of the $\h$ is 
also possible.  The resulting ability to
distinguish different models of the Higgs sector is illustrated by
detailed studies for two-Higgs-doublet models.
\end{abstract}

\medskip

In extensions of the Standard Model (SM) there are multiple
neutral Higgs bosons. Their masses and
couplings are often dependent upon many parameters; CP-violating
mixing of CP-even with CP-odd neutral Higgs fields is generally possible.
Thus, if Higgs boson(s) exist and are discovered at future colliders, it will 
be extremely important to determine both the magnitude and the CP nature
of their couplings\cite{review,otherwork,ggh}.

It has been shown that the process $e^+e^-\to t \anti t \h$ 
($\h$ is our notation for a generic neutral Higgs boson) at the 
proposed Next Linear Collider (NLC) can provide rather accurate
determinations of the CP-even and CP-odd
$t\anti t\h$ Yukawa couplings and at least a rough
value for the $ZZ\h$ coupling \cite{ggh}. The
$\epem\to Z\h$ process provides a direct measurement of the $ZZ\h$
coupling when analyzed in the missing-mass mode with $Z\to \epem,\mupmum$.
Here, we demonstrate that the accuracy with which
the couplings can be determined using these two processes
will discriminate in a very decisive manner 
between different models for the Higgs boson sector.
By way of illustration, we will consider the
general Two-Higgs-Doublet Model (\thdm), for which the masses and couplings
of the three neutral Higgs bosons are all free parameters.
For simplicity, we focus on the $\h$ that is the lightest of the three
eigenstates, and assume that it is sufficiently light 
compared to the other two that 
$\epem\to t\anti t \h$ is not sensitive to the other Higgs states.

The relevant Feynman rules for the $Z\h$ and $e^+e^-\to t \anti t h$ 
processes can be parameterized as:
\begin{eqnarray}
t\anti t h: -\anti t (a+ib\gamma_5) t {gm_t\over 2 m_W}\;,\;\;
ZZh: c {gm_Z\over cos(\theta_W)}g_{\mu\nu}\;,
\end{eqnarray}
where $g$ is the usual electroweak coupling constant. For the SM,
\begin{eqnarray}
a=1\;,\;\;b=0\;,\;\;c=1\;.
\label{abcsm}
\end{eqnarray}
For the \thdm, the couplings are more complicated.
We have
\begin{eqnarray}
a&=&{R_{2j}\over \sin\beta }\;, b=R_{3j}\cot\beta \;,\;\;
c=R_{1j}\cos\beta 
+R_{2j}\sin\beta \;,
\end{eqnarray}
where $j=1,2,3$ indicates one of the three Higgs mass eigenstates,
$\tanb$ is the ratio of the vacua of the neutral members
of the two Higgs doublets (we assume a type-II \thdm), and 
$R_{ij}$ is a $3\times 3$ orthogonal matrix which specifies the transformation
between the \thdm\ Higgs fields and the Higgs boson
mass eigenstates. We employ the parameterization:
\begin{eqnarray}
R = \left (\begin{array}{lll}
c_1&s_1c_3&s_1s_3\\
-s_1c_2&c_1c_2c_3-s_2s_3&c_1c_2s_3+s_2c_3\\
s_1s_2&-c_1s_2c_3-c_2s_3&-c_1s_2s_3+c_2c_3
\end{array} \right )\;,
\end{eqnarray}
where $s_i=\sin\alpha_i$ and $c_i=\cos\alpha_i$.
Without loss of generality, we identify the lightest Higgs $\h$
with the $j=1$ mass eigenstate. In this case, we have
\begin{equation}
a=-{s_1c_2\over \sin\beta }\;,~~~b= s_1s_2\cot\beta \;,~~~
c = c_1\cos\beta -s_1c_2 \sin\beta \;.
\label{abcsol}
\end{equation}
The $\h$ has CP-violating couplings if either $ab\neq0$ or $bc\neq 0$.

We make a few remarks based on Eq.~(\ref{abcsol})
regarding special limiting cases.
\begin{itemize}
\item
In the \thdm\ context one can always reproduce the SM
couplings of Eq.~(\ref{abcsm}) for any given $\tanb$ by taking
$\alpha_1=\beta$ and $\alpha_2=\pi$. If $\tanb$ is not large then
determination of $\tanb$ would be possible via observation of one of the
other Higgs bosons ($j=2,3$).
However, if $\tanb$ is large then $\alpha_1\sim \pi/2$;
coupled with $\alpha_2\sim \pi$, this implies that the remaining Higgs
bosons ($j=2,3$) will have small $t\anti t\h$ couplings, $\propto
1/\tanb$, and would not be easily probed via the $t\anti t\h$ final state.
\item
At large $\tanb$, $b\sim 0$ and $a\sim c\sim -s_1c_2$ (unless $s_1c_2\to 0$ 
as in the SM limit);
sensitivity to the exact value of $\tanb$ is lost and 
$\alpha_1$ and $\alpha_2$ cannot be independently determined.
\item
For very large $\tanb\sim \mt/\mb$, the $b\anti b\h$ couplings can
be large, in which case the $\epem\to b\anti b\h$ rate would be significant
and could be analyzed using the procedures to be discussed here
for the $t\anti t\h$ final state.
\end{itemize}
In this report, we will focus on models with $\tanb$ in the vicinity of 1
in order to display the full potential of the $t\anti t\h$ final state.

The $\epem\to Z\h$ differential cross section is kinematically trivial.
The only useful observable is the total cross section, $\sigmazh$,
which is proportional to $c^2$.
The differential $\epem\to t\anti t h$ cross section, without measuring 
the polarizations of the fermions, contains five distinct terms:
${d\sigma\over d\phi}=\sum_{i=1}^5 c_if_i(\phi)$, where 
\begin{equation}
c_1=a^2\,;~~c_2=b^2\,;~~c_3=bc\,;~~c_4=c^2\,;~~c_5=ac\,,
\label{cdefs}
\end{equation}
and the $f_i(\phi)$ are theoretically known functions of 
the Higgs mass $\mh$, the machine energy $\rts$, and the 
final state phase space variables, $\phi$, but are not dependent
on the model. Note the absence of any term proportional to $ab$.
The $c_3=bc$ term is explicitly CP-violating.
The total cross section, $\sigmat$, is a particular
linear superposition of the $c_i$:
\begin{eqnarray}
\sigmat = \int {d\sigma\over d\phi} d\phi = 
c_1 g_1 +c_2g_2 +c_4g_4+c_5g_5\;,
\label{sigt}
\end{eqnarray}
where the $g_i$ are functions of $\mh$ and $\rts$, and
specific experimental cuts. There is no contribution from
the CP-violating $c_3=bc$ component of $d\sigma/d\phi$.
Since many different Higgs 
parameter choices can yield any given value of $\sigmat$,
it is vital to make use of the much greater information
embodied in the detailed dependence of $d\sigma/d\phi$ on the $\phi$
variables.

The statistically optimal technique for extracting the $c_i$ using
$d\sigma/d\phi$ was developed in Ref.~\cite{ggh}. One employs
weighting functions $w_i(\phi)$ such that
$\int w_i(\phi)[d\sigma/d\phi]=c_i$, where the $w_i(\phi)$ are uniquely
defined by demanding that the statistical error in the determination
of the $c_i$ is minimized; this is the choice such that the entire
covariance matrix is at a stationary point with respect to varying
the functional forms for the $w_i(\phi)$ while maintaining $\int
w_i(\phi)f_j(\phi)d\phi=\delta_{ij}$. 
The weighting functions are given in Ref.~\cite{ggh}. By employing them,
one finds
\begin{equation}
c_i=\sum_k X_{ik} I_k=\sum_k M_{ik}^{-1}I_k\,,\quad{\rm where}~~
I_k\equiv \int f_k(\phi) d\phi\,,
\label{ikdef}
\end{equation}
with
\begin{equation}
M_{ik}\equiv \int {f_i(\phi)f_k(\phi)\over [d\sigma/ d\phi]} d\phi\,.
\label{mikdef}
\end{equation}

If there are experimental cuts that exclude a portion
of the phase space in $\phi$, they should be included in computing
$M_{ik}$ via the $\int d\phi$ appearing in Eq.~(\ref{mikdef})
in order that optimal statistics be achieved in the presence of
the cuts.  Since the cuts that will be employed are detector dependent
and cannot be determined at this time,
we have opted to compute $M_{ik}$ in the examples to follow without
including any cuts.  However, we {\em will} reduce the total event
rate by an overall efficiency factor, the magnitude for which will be chosen
so as to reflect a reduction due to cuts.

The covariance matrix corresponding to $M_{ik}$ of Eq.~(\ref{mikdef}) is
\begin{equation}
V_{ij}\equiv \langle \Delta c_i\Delta c_j\rangle= 
{ M_{ij}^{-1} \sigmat\over N(t\anti t\h)}\,,
\label{cerror}
\end{equation}
where $N(t\anti t\h)=\lefft\sigmat$ 
is the total number of events, with $\lefft$ being the effective luminosity:
$\lefft=\epst\ltot$, where $\ltot$ is the total 
integrated luminosity and $\epst$ is the efficiency, including
branching ratios for the $t\anti t\h$ to decay into the useful final states.
Since identification of the $t$ and $\anti t$ requires that one
decay semi-leptonically and the other hadronically, $\epst\leq 2B(t\to l\nu
b)B(t\to 2j b)\sim 0.44$. Depending upon how the $\h$ decays, there
may be a further loss for focusing on reconstructable $\h$
final state decays.  There will also be cuts and detector efficiencies.
We adopt the value of $\epst=0.1$.
For the $\mh=100\gev$ value that we shall focus on,
for which $B(\h\to b\anti b)\sim 0.9$ is likely, this is fairly conservative.

In order to compute the expected experimental errors
for the $c_i$, we first compute $M_{ik}$ [using Monte Carlo integration
in Eq.~(\ref{mikdef}) without cuts] and thence, via Eq.~(\ref{cerror}),
the covariance matrix $V$ for the given input model. The confidence
level with which one can rule out parameter choices 
different from those of the input model is then determined
by the associated $\chi^2$ value:
\begin{equation}
\chi^2(t\anti t\h)=\sum_{i,j=1}^5 (c_i-c_i^0)(c_j-c_j^0)V^{-1}_{ij}\,, 
\label{chisqtth}
\end{equation}
with
\begin{equation}
V^{-1}_{ij}={M_{ij} N(t\anti t\h)\over \sigmat}\,.
\label{vinv}
\end{equation}
In Eq.~(\ref{chisqtth}), 
the $c_i^0$ are the values for the input model and the $c_i$ are
functions of the location in $a,b,c$ parameter space of the alternative model,
see Eq.~(\ref{cdefs}). Sensitivity of $\chisq(t\anti t\h)$
to the $a,b,c$ parameters is thus directly determined by the covariance
matrix for a given model.  Typically, one finds that sensitivity
to $c_1=a^2$ is largest, while the weakest sensitivity is to $c_3=bc$.

We note that $\chi^2(t\anti t\h)$ implicitly includes a contribution due 
to the difference in $\sigmat$ for the input model 
as compared to the alternative models.  
In what follows, we shall be implicitly assuming that the only errors
in $\sigmat$ are the statistical ones as incorporated in
$\chi^2(t\anti t\h)$ in Eq.~(\ref{chisqtth}). However, we note
that $\sigmat$ will be subject to systematic error as well.
The main uncertainties arise
from the fact that one must observe $t\anti t\h$ production in one or more
particular final states, leading to uncertainty 
in $\sigmat$ to the extent that the $t$ and $\h$
branching ratios and/or the detection efficiencies
for these particular final state(s) are uncertain.
Thus, we will be implicitly assuming 
that these uncertainties can be kept below the level of the simple
statistical uncertainty.

The statistical analysis for the $\epem\to Z\h$ process is completely
straightforward. A direct (\ie\ independent of Higgs branching ratios)
measurement of $c_4=c^2$ is obtained when the $\h$ is isolated via a peak in 
the $[(p_{e^+}+p_{e^-}-p_Z)^2]^{1/2}$
missing mass distribution, where we require $Z\to \epem,\mupmum$ 
in order to be assured of
the cleanest possible analysis and most reliable absolute normalization.  
The number of $Z\h$ events is given by $N(Z\h)=\leffzh\sigmazh$,
where $\leffzh=\epszh \ltot$, with $\epszh$ being the efficiency
for detecting the events using $Z\to\epem,\mupmum$ decays:
$\epszh= B(Z\to\epem,\mupmum)\hat\epszh$.
We take $\hat\epszh=0.5$ for the remnant efficiency associated with
cuts and overall detector efficiencies.
The relative accuracy of the measurement of $c_4$ is 
simply given by $1/\sqrt{N(Z\h)}$,
and thus the $\chi^2$ associated with choosing a value of $c_4$ that
differs from that of the input model value $c_4^0$ is given by
\begin{equation}
\chi^2(Z\h)={[c_4-c_4^0]^2\over [c_4^0]^2}N(Z\h)\,.
\label{chisqzh}
\end{equation}
The total $\chi^2$ associated with choosing values for $a$, $b$ and $c$
that differ from the input model values is given by
summing the $t\anti t\h$ and $Z\h$ results:
\begin{equation}
\chi^2=\chi^2(t\anti t\h)+\chi^2(Z\h)\,.
\label{chisqtot}
\end{equation}

We now provide several examples. We 
take $\mh = 100\gev$ and $\sqrt{s} = 1\tev$. We assume $\ltot=500\fbi$
(as achieved for 2 1/2 years of running at $L_{\rm year}=200\fbi$).
This gives $\lefft=50\fbi$ and $\leffzh=16.9\fbi$. 
Assuming no cuts, the above $\mh$ and $\rts$ imply 
$\sigmazh=13.6c_4^0\fb$, yielding $N(Z\h)=229.3c_4^0$ events
for a given input model. For $\sigmat$ [see Eq.~(\ref{sigt})]
we find (fb units)
\begin{equation}
g_1=2.70\;,\;\;g_2=0.530\;,\;\;g_4=0.083\;,\;\;g_5=-0.055\;.
\label{gi}
\end{equation}
Note the insensitivity of $\sigmat$ to $ac$, 
and very modest sensitivity to $c^2$.

We consider three input model cases:
\begin{itemize}
\item SM: We assume that the input model is such that the 
Higgs has SM couplings, Eq.~(\ref{abcsm}).
From Eqs.~(\ref{abcsm}), (\ref{sigt}) and (\ref{gi}), we find
$\sigmat=2.73\fb$, yielding $N(t\anti t\h)\sim 136$ for $\lefft=50\fbi$.
For $\leffzh=16.9\fbi$ we obtain $N(Z\h)\sim 229$.
\item \thdm(I): We assume that the input model is the \thdm\ model
with $\tanb=0.5$, $\alpha_1=\pi/4$, and $\alpha_2=\pi/4$,
yielding $a=-1.118$, $b=1$, $c=0.4088$.  
In this case, as compared to SM couplings, 
$\sigmazh$ is smaller, yielding $N(Z\h)\sim 38$,  and
$\sigmat$ is larger, $\sigmat=3.94\fb$, yielding $N(t\anti t\h)\sim 197$.
\item \thdm(II):
We assume that the input model is the \thdm\ model
with $\tanb=0.5$, $\alpha_1=\pi/4$, and $\alpha_2=\pi/2$,
yielding $a=0$, $b=1.414$, $c=0.6325$.  
In this case, as compared to SM couplings, 
$\sigmazh$ is smaller, yielding $N(Z\h)\sim 92$,  and
$\sigmat$ is also smaller, $\sigmat=1.09\fb$, yielding $N(t\anti t\h)\sim 55$.
\end{itemize}
In all cases, we use Monte Carlo integration
to compute $M_{ik}$ as given in Eq.~(\ref{mikdef}), and Eqs.~(\ref{cerror})
and (\ref{vinv}) to compute the matrix $V^{-1}$; all depend
upon the input $c_i^0$.  In computing $M_{ik}$ we do not

It is useful to note that the $1\sigma$
statistical errors (expressed in percentage terms)
in $\sigmat$ and $\sigmazh$, corresponding
to the above-quoted event rates are:
\begin{equation}
\begin{array} {cccc}
 & {\rm SM} & {\rm \thdm(I)} & {\rm \thdm(II)} \\
\sigmat: & \pm 8.6\% & \pm 7.1\% & \pm 13.5\% \\
\sigmazh: & \pm 6.6\% & \pm 16.2\% & \pm 10.4\%
\end{array}
\label{sigmaerrors}
\end{equation}
We believe that the systematic errors in $\sigmat$ and $\sigmazh$
will be smaller than the above numbers given that the detector efficiencies
and the relevant $t$ and $\h$ branching ratios
should be very well known by the time this analysis is performed.
This is presumed to be the case in obtaining the numerical results
that follow.

\begin{figure}[htb]
\leavevmode
\begin{center}
\centerline{\psfig{file=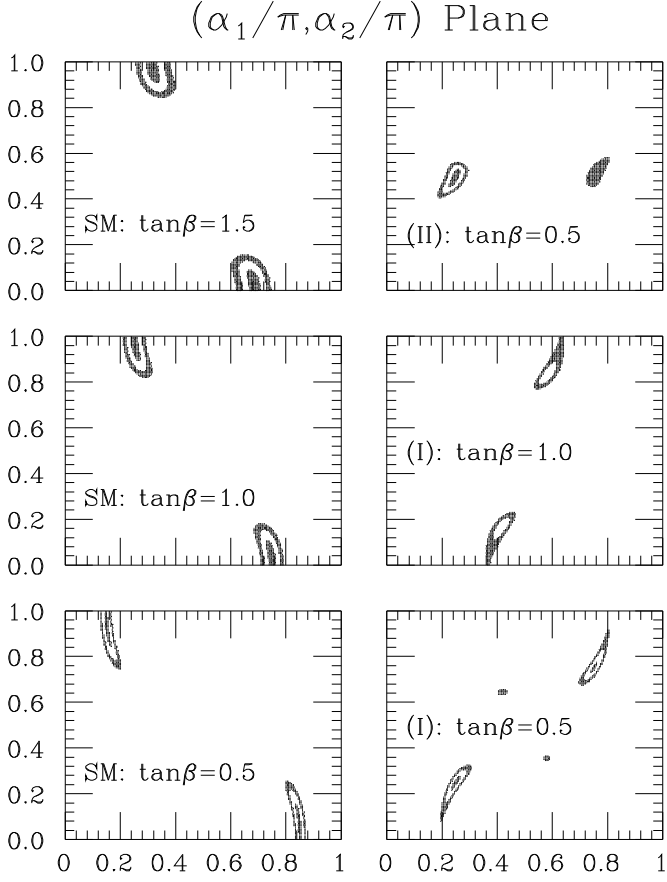,width=3.5in}}
\bigskip
\end{center}
\caption{Regions of $\chisq\leq1$, $1<\chisq\leq 4$ and
$4<\chisq\leq9$ in the $\alpha_1/\pi$ (horizontal axis)
and $\alpha_2/\pi$ (vertical axis) plane. See text for details.}
\label{fig1}
\end{figure}

\begin{figure}[htb]
\leavevmode
\begin{center}
\centerline{\psfig{file=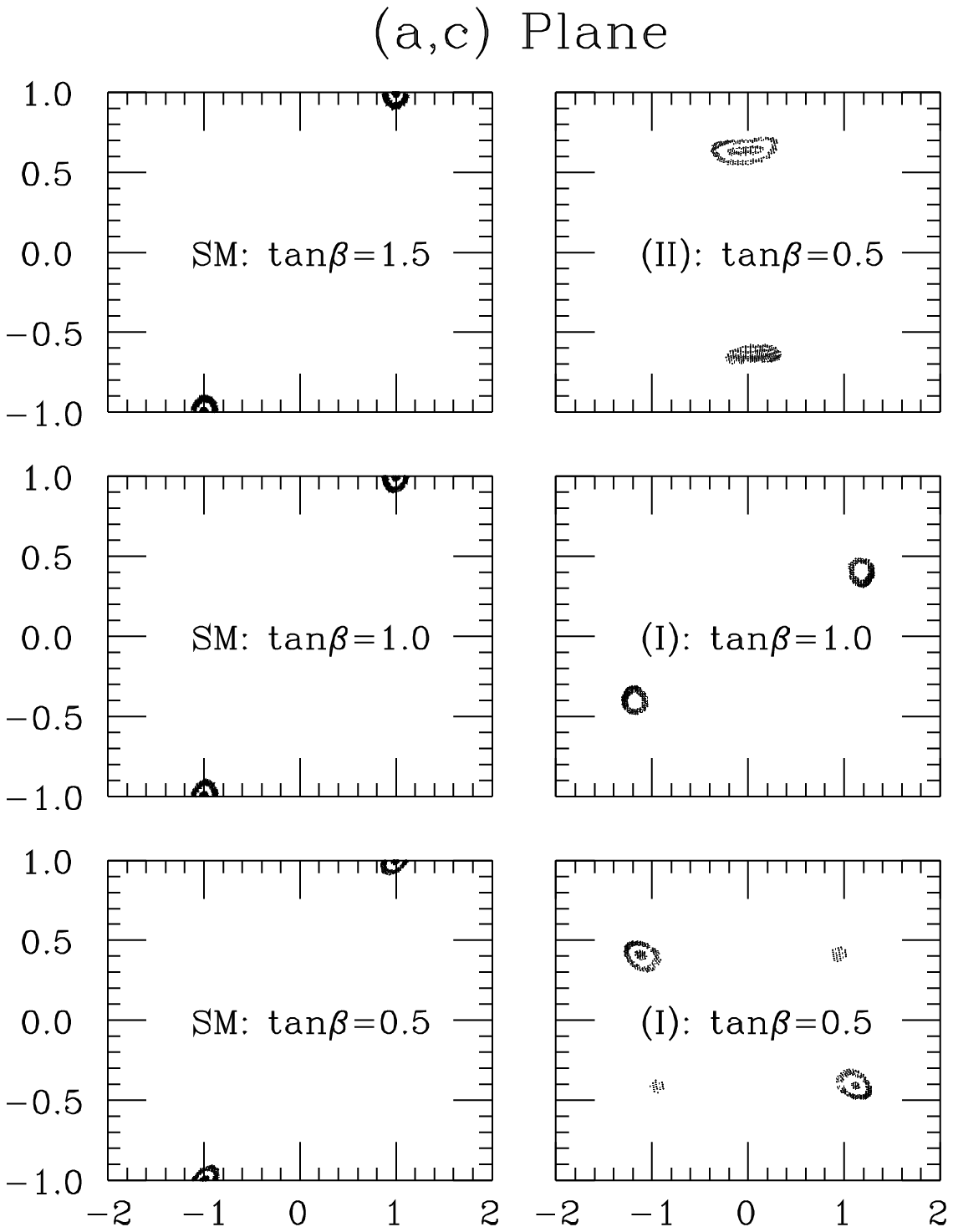,width=3.5in}}
\bigskip
\end{center}
\caption{Regions of $\chisq\leq1$, $1<\chisq\leq 4$ and
$4<\chisq\leq9$ in the $a$ (horizontal axis)
and $c$ (vertical axis) plane. See text for details.}
\label{fig2}
\end{figure}

\begin{figure}[htb]
\leavevmode
\begin{center}
\centerline{\psfig{file=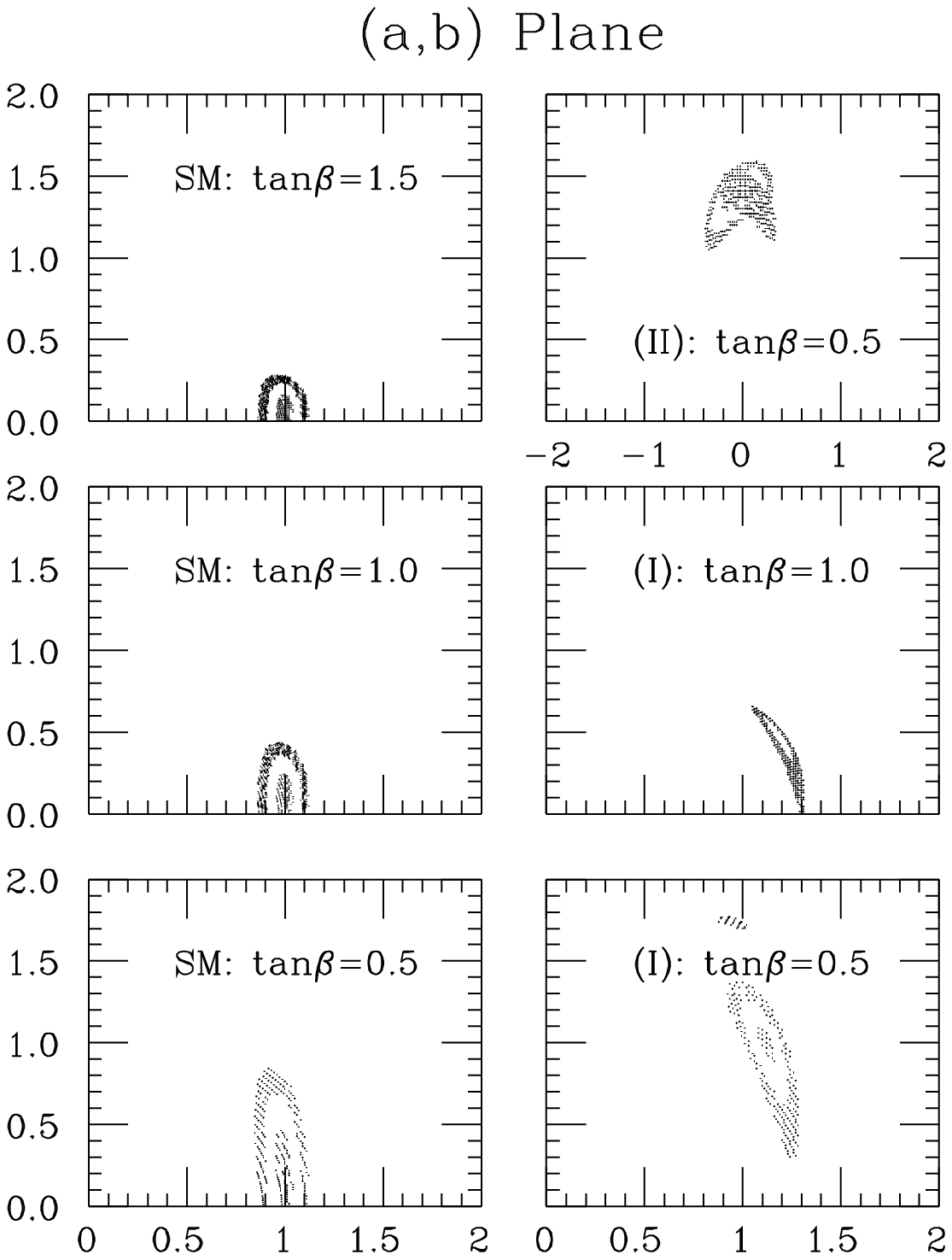,width=3.5in}}
\bigskip
\end{center}
\caption{Regions of $\chisq\leq1$, $1<\chisq\leq 4$ and
$4<\chisq\leq9$ in the $a$ (horizontal axis)
and $b$ (vertical axis) plane. See text for details.}
\label{fig3}
\end{figure}

The accuracy with which the \thdm\ parameters can be determined
is illustrated in Figs.~\ref{fig1}, \ref{fig2} and \ref{fig3}.
Each figure has six windows.  In each window of the
three figures, a filled central region, an empty band, and a filled
band may all be visible. The central region is the $\chisq\leq 1$
region, the empty band is the $1<\chisq\leq 4$ region,
and the outer filled band is the $4<\chisq\leq 9$ region.
If no filled central region is visible, the central region being empty,
then this means that $\chisq\leq 1$ was not possible.  If only a completely
filled region appears, then $\chisq\leq 4$ was not possible.
In the three left-hand windows of each of the three figures,
results are displayed for the case where 
the input model is a \thdm\ constrained so as
to reproduce the SM couplings when $\tanb=0.5$, 1.0 or 1.5.
In the right-hand windows we show results for \thdm(I) with $\tanb=0.5$
and $\tanb=1$ and for \thdm(II) with $\tanb=0.5$.  For all \thdm(I)
[\thdm(II)] parameter choices, $\chisq>9$ if $\tanb=1.5$ [$\tanb=1.0$ or 1.5].

In Fig.~\ref{fig1}, we show the above-described $\chisq$ regions
in $(\alpha_1/\pi,\alpha_2/\pi)$ parameter space. 
We restrict the plot to $0\leq\alpha_1\leq\pi$ and $0\leq\alpha_2\leq\pi$. 
(Since $d\sigma/d\phi(t\anti t\h)$
and $\sigmazh$ are only sensitive to $a^2$, $c^2$, $b^2$, $ac$ and $bc$,
nothing changes if we simultaneously flip the signs
of $a,b,c$. Restricting to $0\leq\alpha_{1,2}\leq\pi$ avoids this ambiguity.)
Note the fact that for all but \thdm(II) the regions
in the $0\leq\alpha_1\leq \pi/2$ domain are self-similar to 
the regions in the $\pi/2\leq\alpha_1\leq 1$ domain obtained by $\alpha_1\to
\pi-\alpha_1$ and $\alpha_2\to \pi-\alpha_2$, which changes the signs
of $a$ and $c$, but not $b$. 

From this figure, we observe the following.
\begin{itemize} 
\item In the case of SM 
input couplings, $\alpha_{1,2}$ must lie close to the $\alpha_2=\pi$
and $\alpha_1=\beta$ values that yield $a=c=1,b=0$, or else to
the $\alpha_2=0$, $\alpha_1=\pi-\beta$ values
that yield $a=c=-1,b=0$, leaving $a^2,c^2,ac$ unchanged.  Note 
that the different $\chisq$ regions all shrink with increasing $\tanb$.
\item In the case of \thdm(I),
only $\tanb=0.5$ (the input value) allows $\chisq<1$, 
and the $\chisq<1$ region
corresponds closely to the input $\alpha_{1,2}$ values
of $\alpha_1=\alpha_2=\pi/4$ or the $a,c$ sign-flipped
$\alpha_1=\alpha_2=3\pi/4$ values. (Self-similarity of the
latter $\chisq$ regions to the former is a consequence of
four facts: i) $bc\sim 0.41$ is not large; ii) sensitivity of $d\sigma/d\phi$
to $c_3=bc$ is weak; iii) $a^2=1.25$ {\em is} large; and iv) sensitivity
of $d\sigma/d\phi$ to $a^2$ is substantial.)
If we allow $1<\chisq\leq 4$, the allowed regions
expand considerably, and for $4<\chisq\leq 9$ there are two more
regions that develop with $\alpha_{1,2}$ values that are very
different from the input values.
Further, we observe that 
$\tanb=1.0$ would be allowed at the $\chisq>1$ level for
yet another region of
$\alpha_{1,2}$ values. Values of $\tanb\geq 1.5$
are excluded at the $\chisq\leq 9$ level.
\item In the case of \thdm(II), only $\tanb=0.5$ (the input value)
allows $\chisq\leq 9$. The $\chisq\leq1$ region corresponds closely
to the input values of $\alpha_1=\pi/4$ and $\alpha_2=\pi/2$.
An alternative region with $\alpha_1\to \pi-\alpha_1$
develops for $4<\chisq\leq 9$. (Self similarity under $\alpha_{1,2}\to
\pi-\alpha_{1,2}$ is not present since, unlike \thdm(I), $a^2=0$
and $bc\sim 0.89$ is fairly large.)
\end{itemize}
Note that in the non-SM \thdm(I) and \thdm(II) 
cases we obtain an approximate determination of $\tanb$.

The implications for the $a,b,c$ couplings appear in Figs.~\ref{fig2}
and \ref{fig3}, in which the $\chisq$ regions are plotted in the $(a,c)$
and $(a,b)$ planes. All regions with $\chisq\leq9$
are shown in the $(a,c)$ plane figure.
In the case of the $(a,b)$ plane, only $a>0$ is shown except
in the \thdm(II) case where both the $a>0$ and $a<0$ regions are shown
(note the difference in horizontal axis labelling for this case).
For the SM and \thdm(I) cases, a self-similar
region to the one displayed for $a>0$ is obtained for 
$a<0$ by flipping about the $a=0$ axis. (Regions with $b<0$
do not emerge.) From the figures, we observe the following.
\begin{itemize}
\item 
For SM input couplings, the output values of $a$ and $c$ must
be very close to the $a=1,c=1$ input values, or the alternative
$a=-1,c=-1$ flip, either of which require $b>0$.
(The $\chisq\leq 1$ regions are very small dots 
in the $(a,c)$ plane; careful examination of the picture is required.)
The value of $b$ is only
moderately well-constrained when $\tanb=0.5$, with $b\leq 0.4$ (0.7)
being allowed at the $\chisq\leq1$ ($\chisq\leq 4$) level. The constraint
on $b$ becomes much tighter as $\tanb$ increases, with $b\leq 0.2$
being required for $\chisq\leq 4$ once $\tanb\geq 1.5$.
\item
For \thdm(I) input, the $\tanb=0.5$ windows of the $(a,c)$ and $(a,b)$
planes show that the $a,b,c$ couplings are all very well-determined
at the $\chisq<1$ level (up to the sign-flip of $a$ and $c$).
Substantial flexibility in $b$ develops for $1<\chisq\leq 4$. 
For $4<\chisq\leq 9$, a region where $a$ has changed sign (but not $c$)
develops.  For $\tanb=1.0$, $\chisq\leq 1$ is not possible, but
for $1<\chisq\leq 4$, a solution develops that has the wrong sign of $ac$
and a very distorted value of $b$. $\chisq\leq9$ is not possible for
$\tanb=1.5$.
\item
For \thdm(II) input, the $a,b,c$ are again well-determined
if we demand $\chisq\leq1$, and $\chisq\leq 4$
allows much less flexibility than
in the \thdm(I) case. However, $4<\chisq\leq 9$ allows a 
a solution with the flipped sign of $ac$ and slightly distorted $b$ values.
[In the $(a,b)$ plane \thdm(II) window, 
the three different $\chisq$ regions associated with
the correct sign of $ac$ are somewhat obscured by
the strange extra blob associated with $4<\chisq\leq 9$ and the wrong
sign of $ac$.]
\end{itemize}

In conclusion, we note that it is very possible (some would say probable) 
that the SM is not correct.  In this case, and if there is a weakly-coupled
Higgs sector, 
there will certainly be Higgs bosons that do not have SM-like couplings.
This is true even if one neutral Higgs is very SM-like 
(as for example is very probable in the minimal supersymmetric model),
since the others must have very small $ZZ$
coupling and can have all manner of $t\anti t$ couplings.
Thus, it will be crucial to determine if an observed Higgs boson
fits into a given model context, such as the two-Higgs-doublet model,
and to determine the model parameters and associated couplings
for acceptable solutions.  By doing this for all the Higgs bosons
we would be able to completely fix the Higgs sector model and parameters.

In this report, we have examined the possibility of carrying
out such a program by applying the optimal
analysis procedure of Ref.~\cite{ggh} to the $\epem\to t\anti t\h$ 
differential cross section and measuring the $\epem\to Z\h$
total cross section. Using $\ltot=500\fbi$
of data from the NLC operating with $\rts=1\tev$, we have demonstrated that 
for models with a reasonable $t\anti t \h$ event rate
the couplings of a 100 GeV \thdm\ Higgs boson 
can be determined with substantial accuracy at the $1\sigma$
level. However, for this luminosity some ambiguities begin to arise
in the $1-2\sigma$ range. Ambiguities at the $\leq 1\sigma$ level
could arise if systematic uncertainties in the 
experimental determination of the overall normalization
of the $t\anti t\h$ and $Z\h$ total cross sections are not small
compared to the statistical accuracies.
At larger Higgs masses, statistics will deteriorate; higher $\ltot$
will be required to avoid significant ambiguity.
However, even when ambiguities emerge, we have found that they are usually
sufficiently limited that the type of analysis presented here will 
make a critical contribution to gaining a clear understanding
of the exact nature of all the Higgs bosons.
Certainly, these procedures will provide a powerful means
for distinguishing between substantially different models. 
We urge our experimental colleagues to carry out fully realistic
simulations of this type of analysis.

\bigskip
\noindent
{\bf Acknowledgement:} Collaboration with B. Grzadkowski in the development
of the optimal technique 
and earlier investigations of the $t\anti t\h$ process (as summarized
in Ref.~\cite{ggh}) is gratefully acknowledged.
%

%

\end{document}